\documentclass[a4paper,11pt,notitlepage]{article}
\usepackage[dvips]{epsfig}
\usepackage{graphics}
\usepackage{rotate}
\font\tenmsb=msbm10
\def\Bbb#1{\hbox{\tenmsb#1}}

\begin{document}

\title{Self-organized criticality \\
in evolutionary systems with local interaction\thanks{%
This work has been supported by the Spanish Ministry of Education, CICYT
Project nos. 96-0168 and 97-0131. We are grateful to Andreu Mas-Colell,
Henrik J. Jensen, and threee anonymous referees for very helpful comments.}}
\author{Alex Arenas \\
Departament d'Enginyeria Inform\`{a}tica, Universitat Rovira i Virgili
(Tarragona, Spain), \and Albert D\'{i}az-Guilera, Conrad J. P\'{e}rez \\
Departament de F\'{i}sica Fonamental, Universitat de Barcelona (Spain), \and %
Fernando Vega-Redondo\thanks{%
Corresponding author -- mail address: Facultad de Econ\'{o}micas,
Universidad de Alicante, 03071 Alicante, Spain; Ph.: 34-96590-3616; Fax:
34-96590-3685; e-mail: vega@merlin.fae.ua.es.} \\
Departamento de Fundamentos del An\'{a}lisis Econ\'{o}mico and \\
Instituto Valenciano de Investigaciones Econ\'{o}micas\\
Universidad de Alicante, Spain\\
and\\
Departament d'Econom\'{i}a i Empresa \\
Universitat Pompeu Fabra, Barcelona, Spain}
\date{First version: May 1999; revised: December 2000}

\maketitle

\begin{abstract}
This paper studies a stylized model of local interaction where agents choose
from an ever increasing set of vertically ranked actions, e.g. technologies.
The driving forces of the model are infrequent upward shifts (``updates''),
followed by a rapid process of local imitation (``diffusion''). Our main
focus is on the regularities displayed by the long-run distribution of
diffusion waves and their implication on the performance of the system. By
integrating analytical techniques and numerical simulations, we come to the
following two main conclusions. (1) If dis-coordination costs are
sufficiently high, the system behaves \emph{critically}, in the sense
customarily used in physics. (2) The performance of the system is optimal at
the frontier of the critical region. Heuristically, this may be interpreted
as an indication that (performance-sensitive) evolutionary forces induce the
system to be placed ``at the edge of order and chaos''.{\Large \medskip }

\textbf{JEL classif. nos.}: C72, O31

\textbf{Keywords}: Self-organization, criticality, local interaction,
technological change.
\end{abstract}

$
\begin{array}{c}
\,
\end{array}
$\pagebreak

$
\begin{array}{c}
\,
\end{array}
$\bigskip

\bigskip

\section{Introduction}

\setlength{\baselineskip}{20pt}\renewcommand{\baselinestretch}{1.3}Social
and economic change typically comes in ``waves'' or ``avalanches'' with
seemingly little intertemporal structure. Its complex dynamics are partly a
result of the following two features: (a) the stimulus for change spreads
throughout the population by cumulative local interaction channeled along a
social network; (b) the payoff incentives that govern individual behavior
are subject to\ considerations of (local) compatibility and/or coordination
with neighbors. In this paper, we focus on a simple model displaying these
two features. To fix ideas, we shall often propose a technological
interpretation of the model and conceive its dynamics as an unbounded
process of technological change. Its motion may then be seen as fueled by
two forces: (i) infrequent upward perturbations, which are payoff
independent and will be simply referred to as ``updates''; (ii) relatively
rapid payoff-responsive adjustment, which will be know as ``diffusion''.
Thus, for the sake of focus, we study a context where diffusion is the only
\emph{purposeful} activity, reacting to a stochastic process of gradual
``invention'' that is fully \emph{exogenous}.

In our stylized framework, it turns out that different technological
scenarios can be solely characterized by the value of a certain cost related
parameter $k\in \Bbb{R}_{+}.$ This parameter reflects the relative costs of
upward versus lower dis-coordination with neighbors, higher values of $k$
indicating a situation where the former costs are relatively larger than the
latter. In this context, our main conclusions can be summarized as follows.
First, we find that there is a certain threshold for $k$ which (roughly)
separates the region where the population evolves in almost perfect
coordination from the region where a wide range of technological
heterogeneity tends to persist over time. Moreover, within the latter
region, the system behaves in a \emph{critical} manner, the concept of
criticality being understood here as in modern physics (Bak \emph{et al.}
(1987)). In particular, it refers to the fact that the distribution over
``avalanche sizes'' (i.e. the range of diffusion waves) generated by the
recurrent updates obeys a power law. This fact has interesting theoretical
implications. For example, it indicates that the system displays no
characteristic scales (or magnitudes), the trade-off between avalanche size
and its corresponding empirical frequency remaining constant throughout (in
terms of proportional changes).\footnote{%
Such a constant tradeoff can also be understood as an indication that local
stimuli have a relatively large (i.e. non-local)\ range of influence. (For
example, this would not occur if the decay of avalanche frequency with
respect to size were exponential.) Traditionally, it was thought that
physical environments had to be \emph{accurately }tuned to some critical
state in order to obtain such long-range effects. Since the work of Bak
\emph{et al.} (1987), such criticality is known to be attainable without any
external fine-tuning (i.e. through self-organization).} But perhaps more
importantly for our present purposes, another useful implication of
criticality pertains to the extreme simplicity displayed by long-run
aggregate behavior under those circumstances. Such simplicity will render it
tractable to carry an analytical study of the model, thus shedding light on
some important issues. In particular, our main focus here will be on how the
long-run performance of the system depends on $k,$ the key parameter of the
model$.$

In this latter respect, our main conclusion is quite stark: under some
natural measure of ``performance'' (see later for details), the behavior of
the system is optimized within a rather \emph{thin} region for $k$ where the
system has already abandoned the synchronization region (a region that
physicists often call supercritical) and has barely entered the critical
region. Borrowing a well-known phrase of Kauffman (1993), we may interpret
this conclusion as a tangible embodiment of the tenet that many dynamical
systems have their performance optimized at the thin ``edge of order and
chaos''.

The reason why \emph{some} heterogeneity must be conducive to better
performance is not difficult to comprehend: only if some inter-agent
differences arise along the process may diffusion have a significant role to
play. However, to understand why optimality should be attained at the
``brink'' where such heterogeneity is about to recede is a quite more subtle
issue. In essence, we shall find it to be a consequence of the following
intriguing feature displayed by critical behavior: across different
avalanche sizes \emph{and }values of $k$, there is a \emph{constant}
(proportional) effect of avalanche size on the induced technological advance.

In this paper, our approach to studying these phenomena relies on a synergic
use of both numerical simulations and analytical techniques. On the one
hand, we resort to numerical simulations to obtain some regularities (e.g.
the verification of power laws) on which to build our ensuing formal
analysis. This analysis is then used to gain a theoretical understanding of
some of the key conclusions also obtained from numerical simulations (e.g.
those concerning the optimal performance of the system). Overall, it is such
a ``symbiosis'' of mathematical analysis and numerical results that will
provide us with most of the insights obtained on a dynamical system that,
because of its huge complexity, would be hardly tractable otherwise (e.g. by
an exclusive use of analytical techniques).

This paper owes much of its original inspiration to the booming literature
on self-organized criticality developed in physics over the last decade.
(The seminal articles are Bak \emph{et al.} (1987) and Bak \emph{et al.}
(1988), whereas a good recent survey can be found in the monograph by Jensen
(1998).) This field of research started with the study of simple \emph{%
sandpile-like} setups, where many of the essential ideas and insights first
originated. Subsequently, there has been a wide variety of different
contexts (biological evolution, the time pattern of earthquakes, the
functioning of the brain, the dynamics of traffic jams, etc.) where simple
adjustment rules have been seen to generate self-organized critical behavior
(see Bak (1996) for an informal survey of applications). More in line with
our present concerns, critical behavior has also been observed in the way
``information'' diffuses through social systems. For example, Redner (1998)
has showed that the ``waves'' of academic quotation display power
distributions, and similar conclusions have been found for how connections
are distributed in the internet (Faloutsos \emph{et al. }(1999)) or in the
World-Wide Web (Albert \emph{et al. }(1999) or Huberman and Adamic (1999)).
In economics and game theory, these ideas have received only little
attention. However, they have been applied to the study of business cycles
by Scheinkman and Woodford (1994), economic geography by Krugman (1996), or
games displaying strategic substitutabilities (i.e. ``anti-coordination''
games) by Agliardi (1998, Chapter 6). Our essential change of focus over
this work is that our objective here is not only positive (describing
long-run regularities of complex dynamic systems) but also normative, i.e.
linking the particular form of these regularities to the performance of the
system. In spirit, therefore, our perspective is quite akin to that held by
those researchers (see Arthur \emph{et al.} (1997)\emph{\ }for a wide
collection of representative work) whose objective has been to understand
the interplay between complexity and performance in large socio-economic
systems. This is also the concern of a previous paper of ours (Arenas \emph{%
et al. }(2000)), where some preliminary results along these lines were
originally reported.

Our model has also borrowed some important features from the recent
evolutionary literature on learning in games, initiated by the seminal
papers of Kandori, Mailath \& Rob (1993) and Young (1993). In particular, we
pursue the same methodology of integrating selection adjustment (here,
myopic best response) with occasional perturbations. More specifically, the
present model is akin to those of Blume (1993), Ellison (1993), or Young
(1998) where, as in our case, a topology of local interaction is introduced
and each agent is constrained to playing the game with her immediate
neighbours. The game being played, on the other hand, bears some key
similarities to the so-called minimum-effort game (see Bryant (1983), van
Huyck, Battalio \& Beil (1990) or Crawford (1991)), which has also been
widely studied in the equilibrium-selection literature. Our approach,
however, displays two essential differences with this literature: (i) the
setup involves a ``changing game'' with a unbounded set of potential
actions; (ii) the concern is not one of equilibrium selection but of dynamic
performance.

In fact, it is precisely the ever-changing nature of the (coordination) game
that allows us to interpret the dynamics induced as a process of
technological change carried out under local complementarities (recall our
former discussion). This then brings to mind two vast strands of related
research: the literature concerned with so-called network externalities and
that studying technological change and growth. We address each of them in
turn.

Network externalities have been a major field of study in the theory of
industrial organization at least since Farrell and Saloner (1985) and Katz
and Shapiro (1985) wrote their seminal papers on the subject -- see
Economides (1996) for a recent survey. The primary aim of this literature
has been to understand, from a strategic viewpoint, the considerations that
might promote or deter technological adoption when the payoffs to it are
directly dependent on the decisions of others; specifically, that is, on the
number of other producers (or consumers) producing (or buying) the new good.

On the other hand, concerning the relationship between technological change
and economic growth, this issue has regained a central role in recent times
due to the importance accorded to it by the so-called New Growth Theory. A
good case in point is the Schumpeterian model of growth through ``creative
destruction'' developed by Aghion and Howitt (1992), where innovations (that
improve the whole production activities in the economy) arrive
stochastically through investment in R\&D geared towards the enjoyment of
(temporary) monopoly gains. This model was enriched by Grossman and Helpman
(1991) by integrating the technological-ladder approach proposed by Aghion
and Howitt with the multi-sectorial features of a model formulated by
Sergestrom, Anant and Dinopoulos (1990). In such an enriched Schumpeterian
model, technological ladders are climbed in each separate sector by specific
R\&D\ expenditures targeted to each of them. In every sector, therefore, a
process of creative destruction unfolds, which forever pushes the economy's
average technological level upwards over time.

Heuristically, one could interpret our model as reflecting a stylized
process of technological change that merges the Schumpeterian features
displayed by the Grossman-Helpman approach with the technological
complementarities studied by the aforementioned literature on network
externalities. In contrast with the former, innovation is taken here to be
purely random (i.e. not the outcome of purposeful and forward-looking
agents). And, in contrast with the latter, the technological
complementarities are studied dynamically, thus amounting to an unbounded
process of technological growth through (purposeful, albeit myopic)
diffusion.

The rest of the paper is organized as follows. Section 2 describes the
framework. Section 3 presents the dynamics. Section 4 presents the numerical
simulations: while its Subsection 4.1 focuses on the regularities displayed
by the induced limit distributions, Subsection 4.2 is concerned with
identifying interesting conclusions regarding issues of self-organization
and long-run performance. Section 5 pursues a task of synthesis. That is, it
builds upon the regularities found in Subsection 4.1 to provide an
analytical explanation for the results obtained in Subsection 4.2. Finally,
Section 6 closes the paper with a summary and the discussion of some issues
left for future research.

\section{The framework\label{Fk}}

We consider $n$ agents, each of them occupying a particular node in a
one-dimensional boundariless lattice (i.e. a ring). Time is discrete. At
every $t=0,1,2,...$, each agent $i\in N\equiv \{1,2,...,n\}$ adopts a
certain action $a_{i}(t)\in \Bbb{R}_{+}$ that, for concreteness, may be
conceived as the technology level she currently uses. Every agent $i$ is
assumed to interact with the individual to the right and to the left of her.
For simplicity, these neighbors are taken to be those agents with adjacent
indices, $i+1$ and $i-1,$ where $0$ and $n+1$ are respectively interpreted
as $n$ and $1$. Out of each of her two interactions, agent $i$ obtains
corresponding payoffs, $\psi (a_{i}(t),a_{i+1}(t))$ and $\psi
(a_{i}(t),a_{i-1}(t))$, where $\psi :\Bbb{R}_{+}\times \Bbb{R}%
_{+}\rightarrow \Bbb{R}$ is a fixed and common \emph{payoff function.}

We shall postulate that the payoff function may be written in the following
way:
\begin{equation}
\psi (a,a^{\prime })=f(a)-g(a,a^{\prime })  \label{payoffs}
\end{equation}
for some function $f:\Bbb{R}_{+}\rightarrow \Bbb{R}_{+}$ which is
unboundedly increasing (i.e. $\lim_{a\rightarrow \infty }f(a)=\infty )$ and
a function $g:\Bbb{R}_{+}\times \Bbb{R}_{+}\rightarrow \Bbb{R}_{+}$ which is
bounded and satisfies:
\begin{equation}
g(a,a^{\prime })>0\Leftrightarrow a\neq a^{\prime }.  \label{incomp}
\end{equation}
For any given action $a\in \Bbb{R}_{+},$ $f(a)$ may be viewed as the payoff
ceiling for this action. In view of (\ref{payoffs})-(\ref{incomp}), this
maximum payoff is attained in any given interaction with someone playing $%
a^{\prime }$ if, and only if, $a=a^{\prime }.$ Otherwise, there are some
bounded ``incompatibility costs'' that detract from the base payoff. By way
of illustration, we may think of every two neighboring agents as involved in
the completion of a certain joint project, for which dissimilarity of
actions (or technological levels) leads to some waste of resources.

In general, the incompatibility costs incurred by any given agent may arise
from two alternative sources: (i) the agent is too advanced relative to her
neighbors; or (ii) she is too backwards. In either case, the induced effects
may be of different significance. To account for this possibility, we posit:
\begin{equation}
g(a,a^{\prime })=\left\{
\begin{array}{cc}
\gamma _{1}(a^{\prime }-a)\quad & \mbox{if }a^{\prime }>a \\
\gamma _{2}(a-a^{\prime })\quad & \mbox{if }a>a^{\prime }
\end{array}
\right.  \label{incomp2}
\end{equation}
where the functions $\gamma _{1},\gamma _{2}:\Bbb{R}_{++}\rightarrow \Bbb{R}%
_{++}$ reflect, respectively, the negative payoff consequences of being more
or less (technologically) advanced than one's partner. For simplicity, these
two functions will be taken to be ``scaled symmetric counterparts'' in the
following sense: there are positive parameters $k_{1}$ and $k_{2}$ such
that, for all $x\in \Bbb{R}_{++},$
\begin{equation}
\frac{1}{k_{1}}\gamma _{1}(x)\equiv \frac{1}{k_{2}}\gamma _{2}(x).
\label{incomp3}
\end{equation}
Intuitively, the larger (smaller) is $k_{1}$ as compared to $k_{2}$ the more
(less) detrimental it is to be more advanced than one's partner as opposed
to being more backwards. As it turns out, only the difference $k\equiv
k_{1}-k_{2}$ will play a relevant role in the analysis.

Thus, heuristically, one may conceive the context described as reflecting a
situation where every two neighboring agents, $i$ and $i+1,$ are involved in
a coordination game that, if faced in isolation, would induce both of them
to choosing the same action. Nevertheless, the key feature of our approach
is that every player $i$ must choose a \emph{common} action in each of the
games she plays. Therefore, the games played by agent $i$ with $i-1$ and $%
i+1\ $are not independent, i.e. cannot be treated in isolation. This is
precisely the assumption that renders the model interesting, and is akin to
that posited by the received evolutionary literature concerned with
equilibrium selection in coordination games (recall the Introduction).
Unlike this literature, however, we contemplate an unbounded ladder of
possible actions where players may coordinate. This provides the basis for
the rich adjustment dynamics that will be seen to arise along the induced
``ever-moving game''. A formal description of this dynamics is undertaken in
the next subsection.

\section{The dynamics\label{Dyn}}

Within the basic framework just introduced, we posit an adjustment dynamics
displaying the following two components:

\begin{enumerate}
\item[\emph{Diffusion:}]  When receiving a revision opportunity, each agent $%
i$ behaves ``myopically'' and adopts an action that is a best response to
what her neighbors are currently doing. Such a diffusion component of the
process is taken to operate in a relatively fast manner.

\item[\emph{Updates:}]  Occasionally, an agent is subject to an exogenous
perturbation (that may be conceived as an innovation) which shifts upward
her action (technology level) by some randomly chosen amount. This component
of the process is to be thought as relatively slow.
\end{enumerate}

For the sake of tractability, the diffusion and update processes are
formally decoupled. That is, we postulate that the updates only perturb
(``punctuate'') the system when the diffusion has reached a standstill.
Thus, in between any two consecutive updates, the process is supposed to
have enough time to reach a point where no agent wants to revise her action
any further. This is a convenient but extreme assumption made by the whole
literature on self-organized criticality (recall the Introduction), since it
allows for sharp definitions of the key notions of wave or avalanche (see
below). However, the same long-run qualitative behavior of the model would
be observed if the two dynamics (diffusion and updates) were genuinely
integrated but the former happened to be much faster than the latter. In
this sense, therefore, such less extreme situation can be seen as suitably
approximated by a context where both dynamics are formally separated.\medskip

Next, we describe precisely each of those two components of the dynamics:
first the slow updates, then the fast diffusion.\bigskip

{\large (i) }\underline{{\large Updates}}{\Large \medskip }

Updates are indexed by $t=1,2,....$ As explained, after any given update,
there is a diffusion wave that must reach a standstill before the next
update occurs. Denote by $a(t)\equiv \left[ a_{i}(t)\right] _{i=1}^{n}$ the
profile displayed by the population once the diffusion phase triggered by
update $t$ has come to a halt. Then, update $t+1$ is taken to operate as
follows.

A single agent $\iota (t+1)$ is randomly chosen to have her technological
level subject to an upward shift. Specifically, her new technological level
becomes $a_{\iota (t+1)}(t)+\tilde{\sigma}$ where $\tilde{\sigma}$ is a
i.i.d. random variable, distributed on a finite interval $\left[ 0,v\right] $
according to some continuous density $\varphi .$ These updates play the role
of exogenous (i.e. unmodelled) perturbations that shift the technological
level of a particular agent upwards. They may be provided with several
motivations, e.g. payoff shocks or population renewal. However, our
preferred interpretation is that of ``innovation possibilities'' arising in
conjunction with ``optimistic expectations''.

More specifically, we propose to view an update as embodying some new
option/idea received by the agent in question that, when pursued, this agent
\emph{optimistically} expects to be followed by a sufficient upward
adjustment by her neighbors. Notice that in view of the assumption that
every update originates at an equilibrium state, there is a natural \emph{%
asymmetry} between upward and downward changes in this respect. That is,
even if an agent were to receive exogenously the \emph{option} of moving to
a lower action (i.e. a downward update), she will never choose to do so if
she believes that others would only react (if at all) in the same direction.
For, in this case, whatever expectations she might hold on the pattern of
adjustment that could ensue, she can only loose by adopting a downward
update.

As explained below (see (ii)), diffusion adjustments turn out to display the
aforementioned monotonicity property, i.e. they always operate in the same
direction as the update that triggers them. Thus, we may rely on the
considerations just described to provide a heuristic motivation for an
update as the \emph{combination} of an innovation plus optimistic
expectations. In this light, it is natural to assume that, once some
optimistic expectations of this kind are in place, they should exhibit some
persistence. That is, they need not be always downgraded (say, to static
expectations\footnote{%
Static expectations (i.e. the belief that the actions currently adopted by
others will remain in place next period) can be taken to underlie the myopic
best-response dynamics posited below for the diffusion dynamics. Thus, in
this light, an update can be regarded as an infrequent and ``optimistic''
deviation of this state of affairs.}) if the expected response fails to
materialize in the ensuing avalanche. Otherwise, updates would generally
prove to be an insufficient ``fuel'' for \emph{sustained }advance, and our
model would eventually become trapped in an uninteresting deadlock.\footnote{%
For example, in the specific context studied in Section \ref{NumA}, we find
that if the process starts at an homogeneous profile \emph{no }single update
can produce a genuine avalanche (involving more than one agent) when $%
k\equiv k_{1}-k_{2}$ is above the threshold value of $3.164.$ Consequently,
under these conditions, the process would not grow at all if the updates did
not exhibit some persistence.} For convenience, we shall introduce such
persistence into the model quite starkly and simply assume that any player
who undertakes an update towards some $a(t)$ at any $t$ never adjusts her
level below $a(t)$ thereafter. Formally, this is captured by defining, for
each agent $i$ and each $t,$ a lower bound $\varsigma _{i}^{t}$ on her
adjustments that is given by the last former update experienced by this
player. That is, the action level attained at that last update is assumed to
act as a floor on player $i$'s future adjustments (see below for
details).\bigskip

{\large (ii) }\underline{{\large Diffusion}}\medskip

After agent $\iota (t)$ has been perturbed at $t,$ a diffusion (or
adjustment) process ensues. Let us index the stages of this process by $%
q=0,1,2,...$ and let $\alpha _{i}^{t}(q)$ denote the action chosen by any
agent $i$ in stage $q.$ With this notation in hand, the diffusion process
may be described as follows.

At $q=0,$ we have $\alpha _{j}^{t}(0)=a_{j}(t-1)$ for $j\neq \iota (t)$ and $%
\alpha _{\iota (t)}^{t}(0)=a_{\iota (t)}(t-1)+\sigma _{t},$ where $\sigma
_{t}$ is the realization of the random variable $\tilde{\sigma}$ at $t.$
Subsequently, at every $q=1,2,...,$ agents are randomly chosen to revise
their action. Specifically, we postulate that any agent $i\in N$ who
receives an adjustment opportunity at stage $q$ chooses an action $\alpha
_{i}^{t}(q)$ that maximizes (myopically) her payoffs under the following
\emph{double} constraint. On the one hand, she cannot surpass the current
action ceiling prevailing in her own neighborhood. On the other hand, her
new action cannot fall below the floor given by her last own update. The
motivation for the first constraint is that any adjustment by player $i$
which exceeds the maximum action level $\beta _{i}^{t}(q-1)\equiv \max
\,\{\alpha _{i+1}^{t}(q-1),\alpha _{i}^{t}(q-1),\alpha _{i-1}^{t}(q-1)\}$
should be conceived as an ``innovation'' and thus restricted to enter the
system as an update.\footnote{%
Thus, implicitly, we suppose that new (technological) information only
diffuses locally. We believe that this informational restriction can be
substantially relaxed but have not yet conducted any detailed explorations
of different alternatives.} The motivation for the second was already
explained above when introducing our formulation of updates.

Formally, each agent $i\in N$ who receives a revision opportunity at $q$ is
taken to solve the following optimization problem:
\begin{equation}
\begin{array}{l}
\max\limits_{\alpha }\,\left[ \psi (\alpha ,\alpha _{i+1}^{t}(q-1))+\psi
(\alpha ,\alpha _{i-1}^{t}(q-1))\right] \bigskip \\
\quad
\begin{array}{ll}
\mbox{s.t.} & \varsigma _{i}^{t}\leq \alpha \leq \beta _{i}^{t}(q-1).\medskip
\end{array}
\end{array}
\label{optim-problem}
\end{equation}
The above problem determines whether the agent in question might be
interested in revising her former action. Denote by $\Lambda _{i}^{t}(q)$
the set of solutions to the above problem, which is assumed non-empty
(possibly including several actions, of which one may be the status-quo $%
\alpha _{i}^{t}(q-1)$). Furthermore, let $\mathcal{I}^{t}(q)\equiv \{i\in
N:\alpha _{i}^{t}(q-1)\notin \Lambda _{i}^{t}(q)\}$ stand for the set of
agents whose prior action is not optimal. Then, if $\mathcal{I}^{t}(q)\neq
\emptyset ,$ we choose at random one individual $i\in \mathcal{I}^{t}(q)$
and make:

\begin{itemize}
\item[(a)]  $\alpha _{i}^{t}(q)\in \Lambda _{i}^{t}(q)$, any choice in $%
\Lambda _{i}^{t}(q)$ with equal probability;

\item[(b)]  $\alpha _{j}^{t}(q)=\alpha _{j}^{t}(q-1)$ for all other agents $%
j\neq i.$
\end{itemize}

Once (a) and (b) have been implemented, the process enters stage $q+1.$ The
payoff function $\psi (\cdot )$ will be assumed to guarantee the following
two-fold property (see (\ref{payoffs-sim}) for an example):\footnote{%
In general terms, if the ceiling-payoff function $f(\cdot )$ is increasing
and weakly convex (e.g. linear) and the cost functions $\gamma _{i}(\cdot )$
are increasing and strictly concave, this is enough to ensure that the
desired property holds.}

\begin{itemize}
\item[($\dagger $)]  $\alpha _{i}^{t}(q)\geq \alpha _{i}^{t}(q-1)$ for all $%
i $ and $q$;

\item[($\ddagger $)]  there is some finite $\bar{q}$ at which there is no
agent left in a position to revise her action, i.e. $\mathcal{I}^{t}(\bar{q}%
)=\emptyset $.
\end{itemize}

Then, having reached the latter $\bar{q},$ we make $a_{i}(t)=\alpha _{i}^{t}(%
\bar{q}).$

The concatenation of diffusion phases and updates defines the dynamical
system under consideration. For each $t,$ the adjustment process that
restores stability after the corresponding update is called a
(technological) \emph{avalanche}, where we use the usual term coined in the
physics literature for this phenomenon. We shall be interested in
quantifying the size $s(t)$ of each avalanche at $t$ as follows:
\begin{equation}
s(t)\equiv \#\{i:a_{i}(t)\neq a_{i}(t-1)\},  \label{size}
\end{equation}
where $\#\{\cdot \}$ stands for the cardinality of the set in question. We
shall also concern ourselves with the total advance triggered by the
avalanche, as given by:
\begin{equation}
H(t)\equiv \sum_{i=1}^{n}\left[ a_{i}(t)-a_{i}(t-1)\right] .  \label{advance}
\end{equation}
As explained next, our numerical simulations show that these magnitudes, $%
s(\cdot )$ and $H(\cdot ),$ display interesting long-run regularities.

\section{Numerical analysis\label{NumA}}

Given the large dynamic complexity of our model, some of its key features
have been determined through the performance of extensive numerical
simulations. For concreteness, the numerical analysis reported here will
focus on a simple scenario where the conditions required from our general
model are satisfied in a specially transparent manner. Other alternative
specifications consistent with the conditions posited in Sections \ref{Fk}
and \ref{Dyn} have been found to yield the same qualitative conclusions.

Essentially, the two only components of the model which need to be specified
pertain to the payoff function $\psi (\cdot )$ and the update density $%
\varphi (\cdot ).$ Concerning the payoff function, let it be given by:
\begin{equation}
\psi (a,a^{\prime })=a-k_{1}(1-\exp (-\left[ a-a^{\prime }\right]
_{+})-k_{2}(1-\exp (-\left[ a^{\prime }-a\right] _{+})  \label{payoffs-sim}
\end{equation}
for some $k_{1},k_{2}>0,$ where $\left[ x\right] _{+}\equiv \max \,\{0,x\}.$
In terms of the general framework given by (\ref{payoffs}) and (\ref{incomp2}%
), this amounts to making
\begin{eqnarray}
f(a) &=&a  \label{def-f} \\
\frac{\gamma _{1}(x)}{k_{1}} &=&\frac{\gamma _{2}(x)}{k_{2}}=1-\exp (-\left[
x\right] _{+}).  \label{def-gamma_i}
\end{eqnarray}
This is obviously compatible with the required conditions -- in particular, (%
\ref{incomp}) and (\ref{incomp3}). Moreover, under suitable parameter
conditions, (\ref{def-f})-(\ref{def-gamma_i}) turns the bilateral situation
faced by every two neighbors into a coordination game which, locally around
an equilibrium, behaves like the well-known minimum-effort game -- a context
widely discussed in the evolutionary and experimental literature (recall the
Introduction). Specifically, the game then displays the following local
property: if two individuals with ``marginally different'' actions play the
game, the one with the lower one obtains a higher payoff. To see this, note
that if we consider two different actions, $a$ and $a^{\prime },$ with $%
\delta \equiv a^{\prime }-a>0$, the function $h(\cdot )$ defined by $%
h(\delta )\equiv \psi (a^{\prime },a)-\psi (a,a^{\prime })$ has the
following derivative at $\delta =0:$
\[
h^{\prime }(0)=1-(k_{1}-k_{2}).
\]
Thus, if $k_{1}-k_{2}>1$ (which is wholly within the range of interest to be
considered below), we have $\psi (a^{\prime },a)-\psi (a,a^{\prime })<0$
provided that $a^{\prime }$ is marginally larger than $a.$

It is easy to check that the payoff function given in (\ref{payoffs-sim})
guarantees that the diffusion dynamics satisfies ($\dagger $)-($\ddagger $)
above. On the other hand, a further convenient feature following from this
formulation is that the decision problem described in (\ref{optim-problem})
has its optimal solution necessarily lie in the set $\{\alpha
_{i+1}^{t}(q-1),\alpha _{i}^{t}(q-1),\alpha _{i-1}^{t}(q-1)\}.$ That is, the
optimal action must be one of those currently chosen in the neighborhood of
the player in question. Obviously, this will facilitate matters in what
follows by allowing us to handle the optimization problem faced by each
agent in a especially simple manner.

Finally, concerning the stochastic updates, we shall postulate that they are
distributed uniformly and independently on $[0,1]$, i.e.
\begin{equation}
\varphi (x)=1,\quad \forall x\in \lbrack 0,1].  \label{unif}
\end{equation}

Much of our analysis will dwell on the effect of the incompatibility-cost
parameters, $k_{1}$ and $k_{2}$, on the long-run evolution of the process.
In fact, to understand the main issues involved, it turns out that only the
difference $k\equiv $ $k_{1}-k_{2}$ needs to concern us. In view of (\ref
{payoffs-sim}), one can interpret $k$ as the cost difference resulting from
``downwards incompatibility'' (i.e. being too advanced) as compared to that
derived from ``upwards incompatibility'' (i.e. being too backwards).

To gain a heuristic understanding of why this cost difference should be the
key parameter, suppose that the system starts at a synchronized state at
some $t.$ That is, the previous state $[a_{i}(t-1)]_{i=1}^{n}$ satisfies $%
a_{i}(t-1)=\hat{a}$ for some $\hat{a}$ and all $i=1,2,...,n$. Now suppose
that the magnitude of the ensuing $t$th update is $\Delta $ and let $i_{o}$
be the particular agent affected by it. We may then ask: When will this
update again lead the system (once the corresponding diffusion phase has
come to an end) into a new synchronized state at $t$? Of course, this will
occur if the payoff to any agent $i\neq i_{o}$ of adopting $\hat{a}+\Delta $
when at least one of his neighbors has done so is higher than if he were to
remain at level $\hat{a}.$ Thus, if we focus on the only non-trivial case
where one neighbor still adopts $\hat{a},$ the relevant inequality is:
\[
2(\hat{a}+\Delta )-k_{1}(1-e^{-\Delta })>2\hat{a}-k_{2}(1-e^{-\Delta }),
\]
or equivalently:
\begin{equation}
k\equiv k_{1}-k_{2}<k^{\ast }(\Delta )\equiv \frac{2\Delta }{1-e^{-\Delta }}.
\label{critic}
\end{equation}
If (\ref{critic}) holds, the update carried out by $i_{o}$ will lead the
system into a new synchronized state at the \emph{common }level $\hat{a}%
+\Delta $, i.e. $a_{i}(t)=\hat{a}+\Delta $ for all $i=1,2,...,n$. Otherwise,
the update will introduce some ``local heterogeneity'' (i.e.
asynchronization) around $i_{o}.$

In general, as $\Delta $ ranges from $0$ to $1$ (i.e. the support of $\tilde{%
\sigma}$), the value for $k^{\ast }(\Delta )$ increases from $\lim_{\Delta
\downarrow 0}k^{\ast }(\Delta )=2$ to $k^{\ast }(1)=3.164$. Thus, in view of
our preceding discussion, one may expect to find a transition from fully
synchronized behavior (i.e. what physicists often call \emph{supercriticality%
}) for $k\leq 2$ to increasingly more heterogenous behavior as $k$ grows
above this threshold. Indeed, such a transition will be clearly observed in
our simulations, with sharply critical behavior (i.e. power laws) arising
when $k$ approaches the higher threshold given by $k^{\ast }(1).$

Next, we discuss our numerical simulations in some detail. We shall focus in
turn on the following issues:

\begin{itemize}
\item  long-run distributions for the sizes of the avalanches;

\item  long-run distributions for the (technological) advances induced by
avalanches;

\item  long-run relationship between size and advance across different
avalanches.\bigskip
\end{itemize}

\subsection{Power laws}

We have obtained the (empirical) avalanche size distributions for different
values of $k$ and different values of $n,$ the latter being considered in
order to check for possible scale effects. The outcome of these simulations
is summarized in Figures 1-2, where the relationship is depicted in doubly
logarithmic scale.\footnote{%
Each point in the diagram gives the average log-frequency across all sizes
in a corresponding interval. To identify graphically every such interval we
take its mid-point (also in logarithmic scale).}\bigskip\

\begin{figure}
\epsfclipon \epsfxsize=0.5\linewidth \rotate[r]{\epsffile{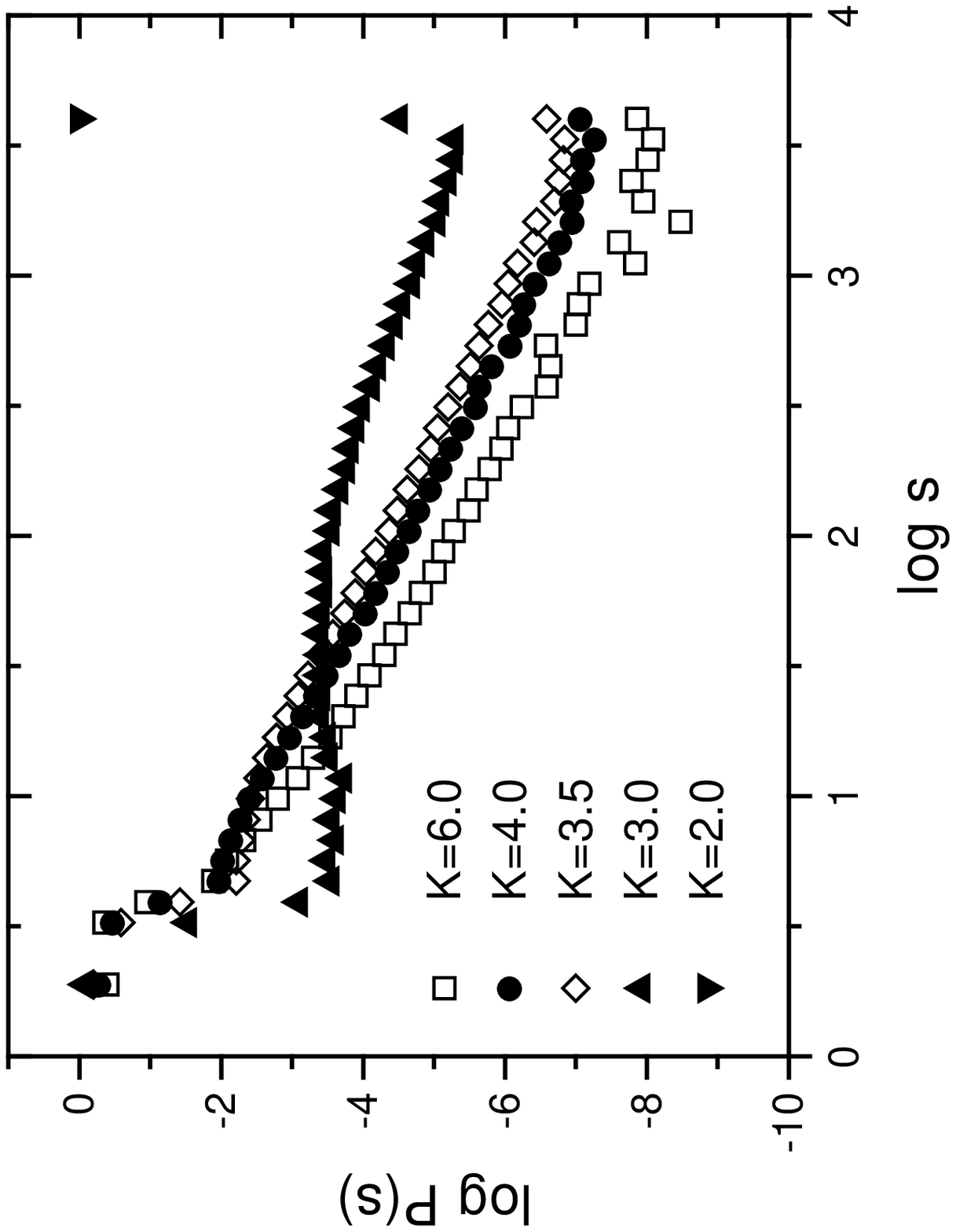}}
\caption{ Distribution of avalanche sizes,
$k=2,3,3.5,4,6;\;n=4096$ -- double logarithmic scale (log
frequency versus log size). }
\end{figure}

\begin{figure}
\epsfclipon \epsfxsize=0.5\linewidth \rotate[r]{\epsffile{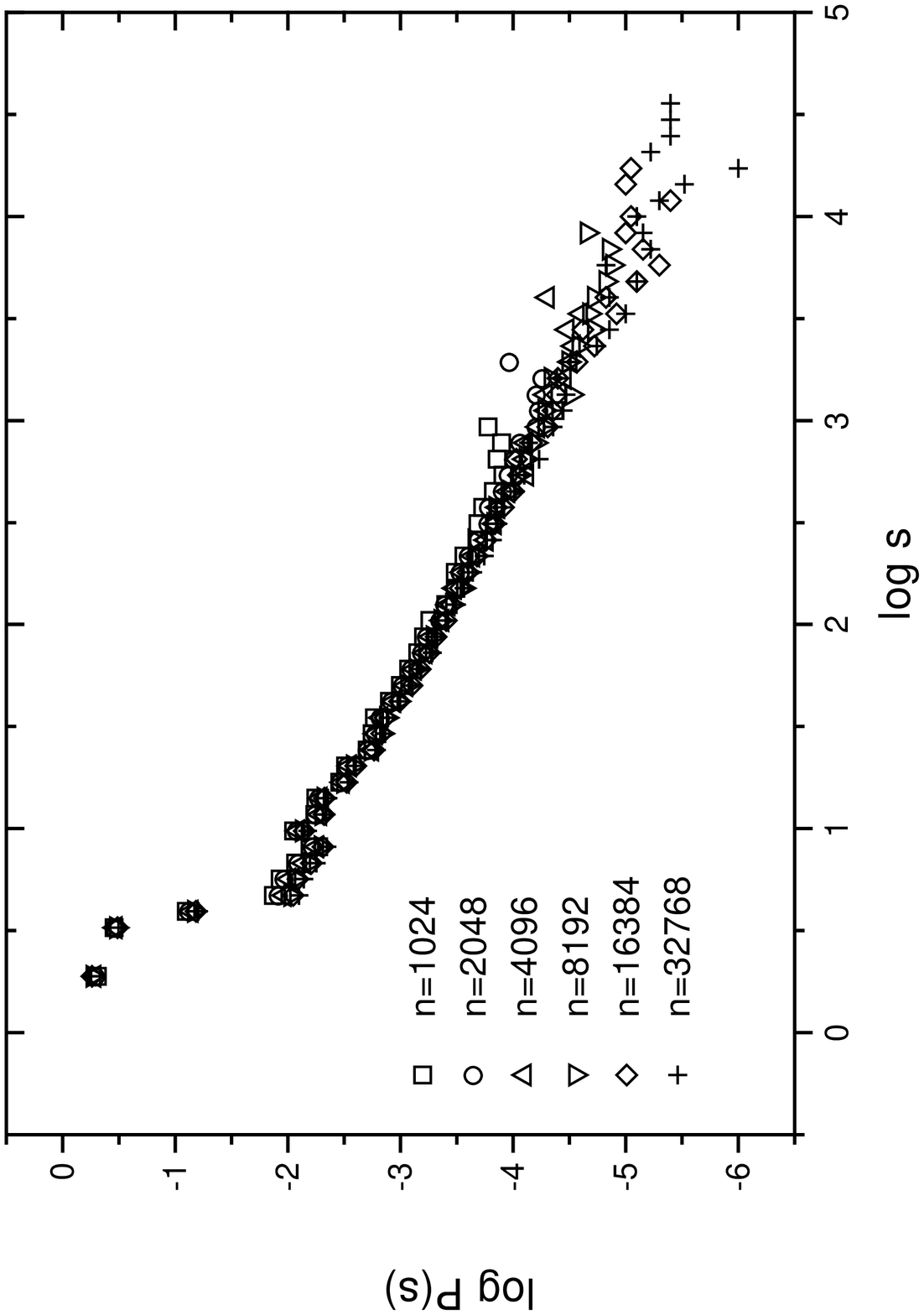}}
\caption{Distribution of avalanche sizes, $k=4;$ $n=1024,$ $2048,$
$4096,$ $%
8192$, $16384,$ $32768$ -- double logarithmic scale (log frequency
versus log size). }
\end{figure}

In Figure 1, we show the size distribution for different values of $k$ and a
fixed population size $n=4096.$ We can notice the transition from a regime ($%
k\leq 2$) where all avalanches are system-size wide to a regime $(k\geq 3.5)$
where the avalanche size sharply obeys a power-law distribution. As this
transition unfolds (e.g. around $k=3),$\footnote{%
For intermediate values of $k\in (2,3),$ the relationships between frequency
and size is found to be increasing, thus reflecting a gradual approach to
the situation where \emph{all }avalances are of full system size as $k$
decreases towards $2$. Figure 1 only shows the results for $k=2,3$ in order
not to complicate the diagram with the depiction of non-critical behavior in
the transition phase.} one observes that avalanches of different sizes do
occur, but they do not yet exhibit the clear-cut power-law regularities that
are the hallmark of criticality.

On the other hand, Figure 2 shows that the power-law relationship between
size and frequency displayed for any given $k$ in the critical region is
independent of population size. For concreteness, this is shown for $k=4$
and $n$ spanning three orders of magnitude, but similar diagrams obtain for
all $k$ in the critical region and even wider ranges in population size.

In Figure 3, we show that analogous conclusions are obtained for the
empirical distribution of advances across different avalanches. This
distribution also obeys a power law for values of $k\geq 3.5,$ which is the
(rough) threshold given before for critical behavior to start manifesting
itself fully in terms of size distributions. This conclusion can be shown to
be independent of population size, much as this was also shown to be the
case in Figure 2 for avalanche size.\bigskip\

\begin{figure}
\epsfclipon
\epsfxsize=0.5\linewidth
\rotate[r]{\epsffile{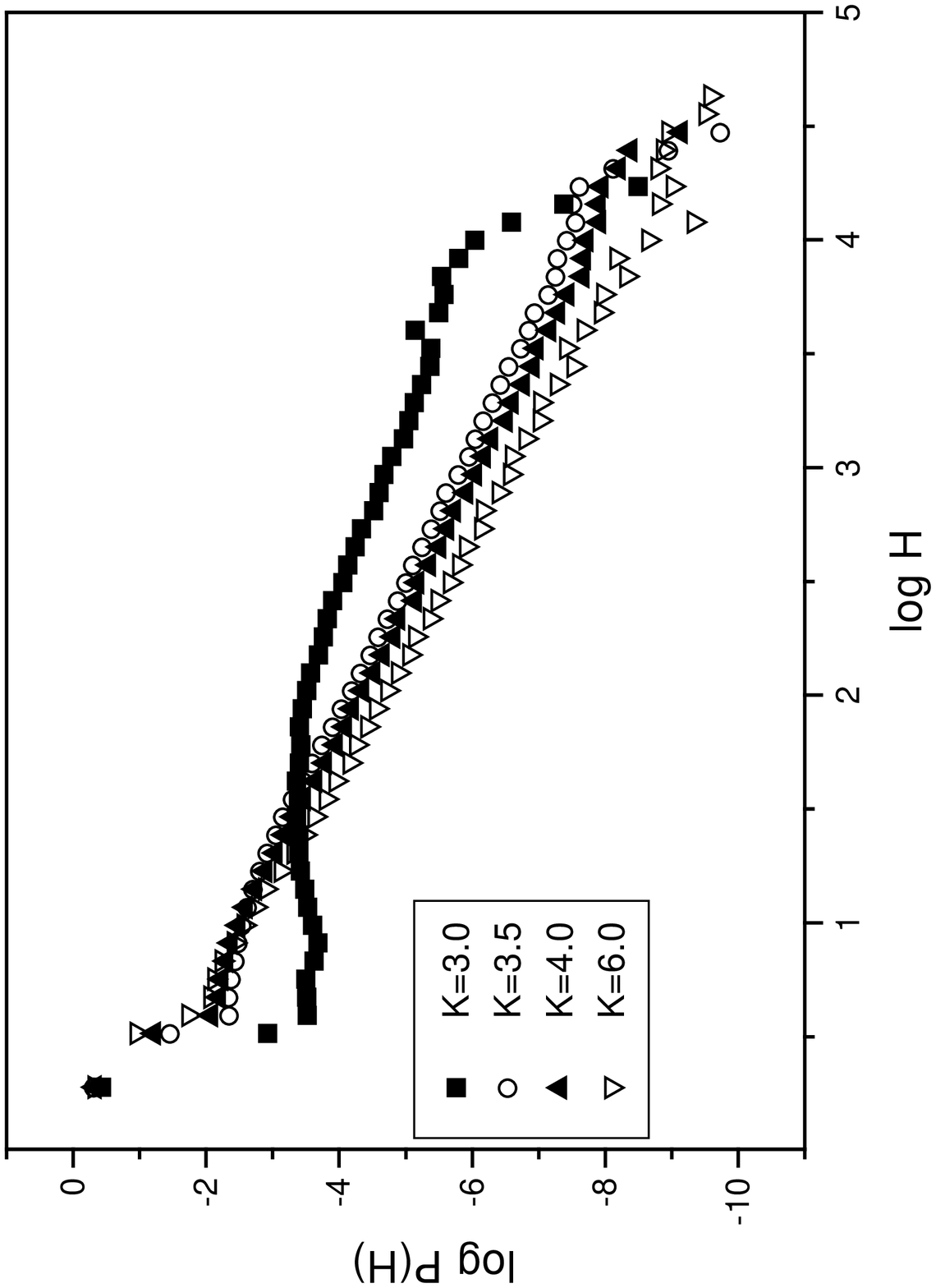}}
\caption{Distribution of avalanche advance, $k=$3, 3.5, 4, 6; $n$=4096 -
double logarithmic (log frequency versus log advance).
}
\end{figure}

The above results indicate that, as soon as downwards (relative)
incompatibility costs become significant, avalanche size and avalanche
advance both display a surprising regularity: the size- or
advance-elasticity of the corresponding long-run frequencies are constant
throughout. In physics, this phenomenon goes by the name of criticality, a
term that points to the absence of characteristic scales, i.e. the lack of
prominent (relative) scales at which the system behavior mainly takes place.

Given the common qualitative features displayed by size and advance
distributions, it is typically expected (see Jensen (1998)) that both
magnitudes also display a power-like relationship between them. Indeed, this
is confirmed by Figure 4, where we have plotted avalanche size versus
associated \emph{average} advance for different values of $k$ in the
critical region ($k\geq 3.5)$.\bigskip\

\begin{figure}
\epsfclipon
\epsfxsize=0.5\linewidth
\rotate[r]{{\epsffile{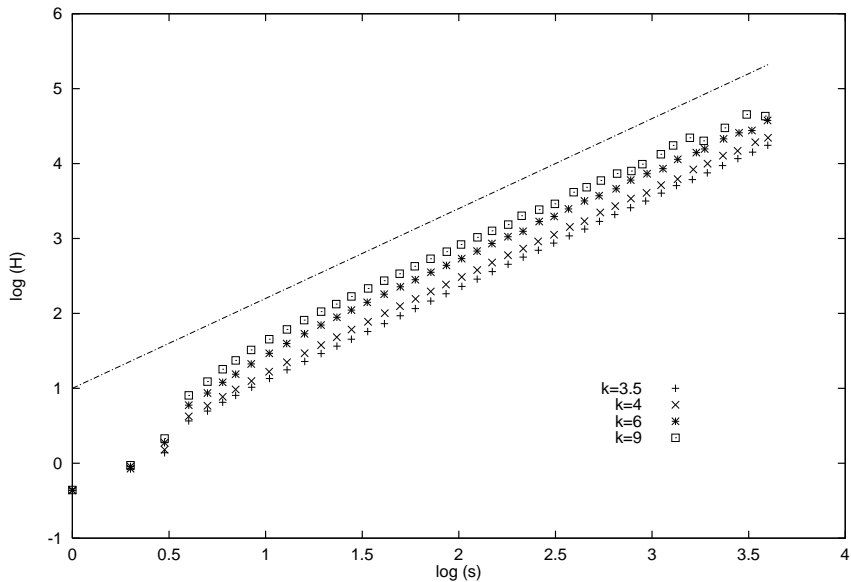}}}
\caption{Average advance versus avalanche size, double logarithmic scale
$k$=3.5, 4, 6, 9; $n$=4096.
}
\end{figure}

Figure 4 also points to an additional important feature of our simulation
results: throughout the critical region$,$ the elasticity of total average
advance with respect to avalanche size (i.e. the slope in double logarithmic
scale) is independent of $k,$ remaining essentially constant to a value
slightly above $1$ (more precisely, at roughly $1.2)$. In contrast with
previous observations (cf. Figures 1-3), note that such a constant
elasticity is not simply a feature that prevails for \emph{given} $k$ but,
rather, is a characteristic that holds uniformly across \emph{different}
(critical) values for $k$. The implications of this intriguing observation
will be discussed below at some length.

Finally, it is worth mentioning that critical behavior is gradually lost for
very large values of $k$.\footnote{%
In our present context, the main distinguishing features of criticality
start to fade away when $k\geq 30.$} Specifically, as $k\rightarrow \infty ,$
any interaction between neighboring sites vanishes and one obtains a process
of so-called random deposition, a well known process in the study of surface
growth (cf. Barabasi \& Stanley (1995)). As usual in these cases, the
transition towards such a state of affairs involves an intermediate phase
where the frequency decay is exponentially decreasing in avalanche size
(i.e. it is much faster than the power decrease displayed in the critical
region). For very large $k,$ therefore, a disproportionate amount of
avalanches are of very small size. Since this state of affairs displays few
implications of interest for our purposes, we do not describe it here in any
detail.

To recapitulate, it is useful to provide a formal summary of the main
power-law regularities observed in our simulations as follows. There are
some thresholds $\underline{k}\,$and $\bar{k}$ defining the critical region
(roughly, $\underline{k}=3.5$ and $\bar{k}=30)$ such that if $\underline{k}%
<k<\bar{k}$:

\begin{enumerate}
\item[\textbf{P1\ }]  The long-run distribution of avalanche sizes $s$
follows a power law of the form:\footnote{%
As standard, the symbol $\sim $ signifies an asymptotic long-run equality
between the two sides of the expression, modulo suitable constants.}
\begin{equation}
P(s)\sim 1/s^{\gamma }  \label{power1}
\end{equation}
for some $\gamma >0,$ dependent of $k$ but independent of population size $%
n. $

\item[\textbf{P2\ }]  The long-run distribution of total advances $H$
follows a power law of the form:
\begin{equation}
P(H)\sim 1/H^{\beta }  \label{power2}
\end{equation}
for some $\beta >0$, dependent of $k$ but independent of $n.$

\item[\textbf{P3\ }]  The relationship between avalanche sizes and
corresponding advances follows:
\begin{equation}
H\sim s^{\alpha }  \label{power3}
\end{equation}
where the exponent $\alpha >1$ is independent of $k$ and $n.$
\end{enumerate}

The above conclusions indicate that, when $k$ belongs to the critical
region, the dynamic behavior of the system is in sharp contrast with that
displayed by the customary approaches to modelling social behavior in
coordination setups. For example, in received models of evolutionary game
theory (both when interaction is assumed local as well as global), a
population facing a coordination game is almost always to be found
``synchronized'' (i.e. at an homogenous equilibrium). Instead, in our case,
the heterogeneous waves (and corresponding advances) realized along the
process will typically induce the rich diversity one often observes in
real-world phenomena (e.g. concerning technological change).

\subsection{Criticality and long-run behavior\label{SS-SOC}}

In view of P1-P3, it is natural to conjecture that some macroscopic
variables (e.g. certain population averages) might exhibit as well
interesting long-run regularities. Indeed, this is confirmed in what follows
concerning two interesting ``summaries'' of the limit behavior of the
system. The first one is a simple measure of population heterogeneity, often
called the width of the system. The second one captures a certain measure of
its performance. We address each of them in turn.

In problems of surface growth in physics, it is common to quantify the
roughness of an interface (what in our case could be conceived as the
heterogeneity of the technological profile) by its so-called \emph{width}.
Restricting attention to states reached once the diffusion process is
complete, the width of the system after update $t$ is defined by:
\begin{equation}
W(t)=\sqrt{\overline{a(t)^{2}}-\overline{a(t)}^{2},}  \label{width}
\end{equation}
where the overline denotes spatial average, i.e.
\[
\overline{a(t)^{q}}\equiv \frac{1}{n}\sum_{i=1}^{n}(a_{i}(t))^{q}.
\]
Of course, the width of the system at any given point in time simply
coincides with the \emph{standard deviation} of the action levels. It is,
therefore, a \emph{global} statistic which would seem to abstract from the
(local) spatial gradients that are the essential driving force of the model.
Our interest in this magnitude, however, derives from the fact that when the
system is critical, its width turns out to be closely related (i.e. is
proportional, in a suitably extended fashion) to the spatial gradient
functions -- see Barabasi \& Stanley (1995) for details. Thus, in this
sense, the width of our critical system provides average ``local''
information and may be regarded as a convenient (indirect) measure of
``technological roughness'' for the induced profiles.

Figures 5 and 6 below depict (again in doubly logarithmic scale) the time
evolution of the width of the system, as defined in (\ref{width}), for two
different values of $k$ in the critical region and several population sizes
(we have found the same qualitative behavior for other values of $k$ in the
critical region and larger population sizes). Since we are only interested
in the \emph{statistical }properties of the profile dynamics, the evolution
of this magnitude is averaged over 1000 independent runs,\footnote{%
Note that, of course, the interface roughness is not a smooth increasing
function of time for each independent run!} the time periods reflecting the
number of updates having materialized up to that point.\bigskip

\begin{figure}
\epsfclipon
\epsfxsize=0.8\linewidth
{\epsffile{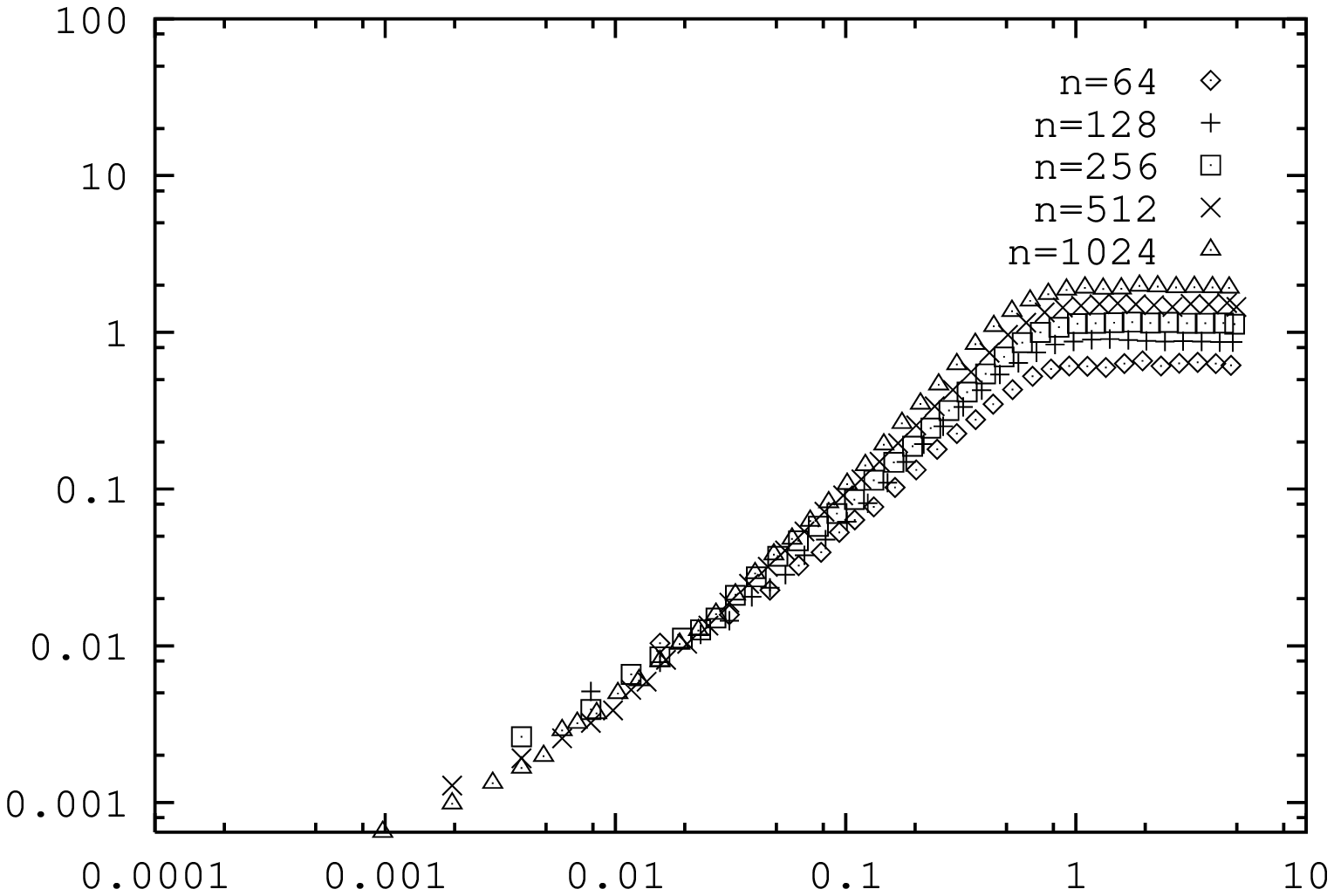}}
\caption{Time evolution of the system width, $k=4;$ $n=64,$ $128,$ $256,$ $
512,$ $1024.$ -- double logarithmic scale (log width versus log (updates)/n).
}
\end{figure}

\begin{figure}
\epsfclipon
\epsfxsize=0.8\linewidth
\epsffile{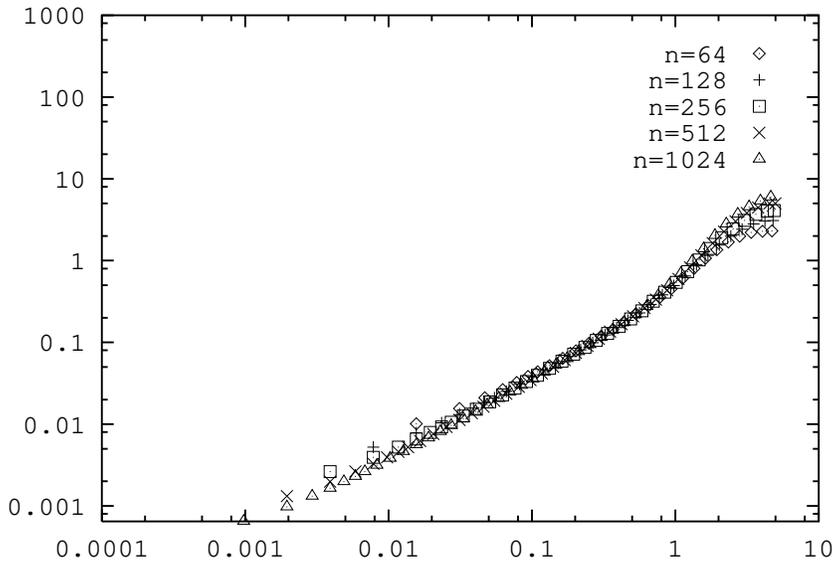}
\caption{Time evolution of the system width, $k=7;$ $n=64,$ $128,$ $256,$ $
512,$ $1024.$ -- double logarithmic scale (log. width versus log. (updates)/n).
}
\end{figure}

The main conclusion to be derived from the above diagrams is that, within
the critical region, there exists a well-defined \emph{long-run level of
heterogeneity} (technological diversity) on which the system tends to settle
as the process evolves and ``self-organizes''. This is another interesting
long-run regularity that follows from critical behavior. It indicates that,
even though avalanches of all sizes (and, therefore, profiles of very
different roughness) are to be observed over time, the long-run level of
heterogeneity (either locally or globally defined) grows monotonically, on
average, to some well-defined magnitude. Naturally, this magnitude depends
on the relevant parameters of the model, $k$ and $n.$ In this respect, the
dependence on $k$ is as one would expect: the larger is $k$, the larger is
the long-run width and the slower (i.e. farther into the future) such a
long-run level is attained. On the other hand, the dependence on $n$ is
equally clear-cut but perhaps less intuitive: population size affects
positively long-run width but has no implications on the speed at which it
is attained.\medskip

Now, we turn to what seems a richer and possibly more interesting
implication of criticality. It concerns the relationship between criticality
and some suitable measure of the performance of the system. To fix ideas,
suppose that the parameter $k$ of our model defines a family of
``technological systems'' (computer architecture, communication protocols,
etc.) that display identical payoff possibilities but differ in their
internal vertical compatibility. More specifically, assume that the payoff
potential for each system grows along a common ``performance ladder'' that
is applicable only when agents coordinate on the same action. On the other
hand, pertaining to dis-coordinated situations, each system differs in the
relative magnitudes of upward versus downward compatibility, as captured by
their corresponding value of $k$ in (\ref{payoffs-sim}).

Suppose that the user population consists of individuals belonging to a
certain organization (say, a big firm) and every adjustment involves a
certain (arbitrary) cost $c>0$ that is independent on the extent of the
change -- for example, any adjustment might involve buying a new piece of
equipment at a fixed cost. In this context, we may ask ourselves the
following question: What is the cheapest way (i.e. cheapest technological
system) by which the organization may eventually attain some pre-specified
(average) technological level? Clearly, if the pre-specified level is high
enough, such a cost minimization is essentially equivalent to a maximization
of the following magnitude:
\begin{equation}
\rho =\lim_{T\rightarrow \infty }\rho (T)=\lim_{T\rightarrow \infty }\frac{%
\sum_{t=1}^{T}H(t)}{\sum_{t=1}^{T}s(t)},  \label{rho}
\end{equation}
where $H(t)$ and $s(t)$ are defined by (\ref{size}) and (\ref{advance}). For
want of a better term, $\rho $ will be called the \emph{performance rate} of
the system.

Consider now a related route to motivate the above measure of performance.
Suppose that, within every time period, there is a fixed amount of resources
that can be devoted to the upgrading the actions (technologies) of the
population, each such adjustment (purchase of new equipment) still requiring
some fixed cost $c>0$. Furthermore, assume that there are always enough
candidates for upgrades (either as ``adjustment'' or ``updates''), but those
that are geared towards matching a neighbor's action (diffusion) always
enjoy higher priority than those that are \emph{not} (i.e. updates). In this
context, maximizing the long-run time \emph{rate }of technological change is
equivalent to maximizing $\rho ,$ as defined in (\ref{rho}).

To facilitate the discussion, write $\rho (T,k)$ and $\rho (k)$ to reflect
the dependence of $\rho (T)$ and $\rho $ on $k.$ Several interesting
conclusions concerning these magnitudes are obtained from our numerical
simulations, as depicted in Figures 7 and 8.\bigskip

\begin{figure}
\epsfclipon
\epsfxsize=0.7\linewidth
\epsffile{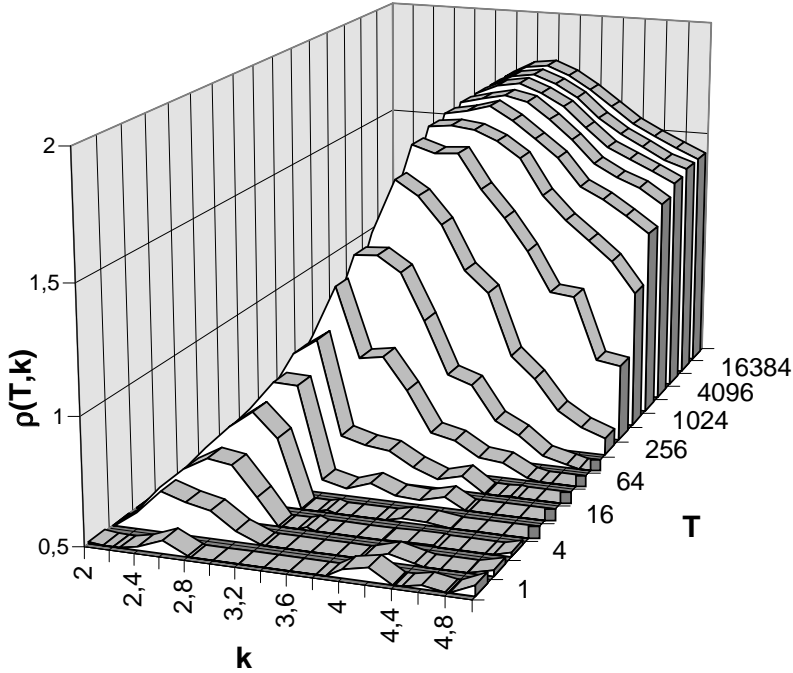}
\caption{Time evolution of performance rate $\rho $ for different values of
$k$ and $n=512.$
}
\end{figure}

\begin{figure}
\epsfclipon
\epsfxsize=0.6\linewidth
\rotate[r]{\epsffile{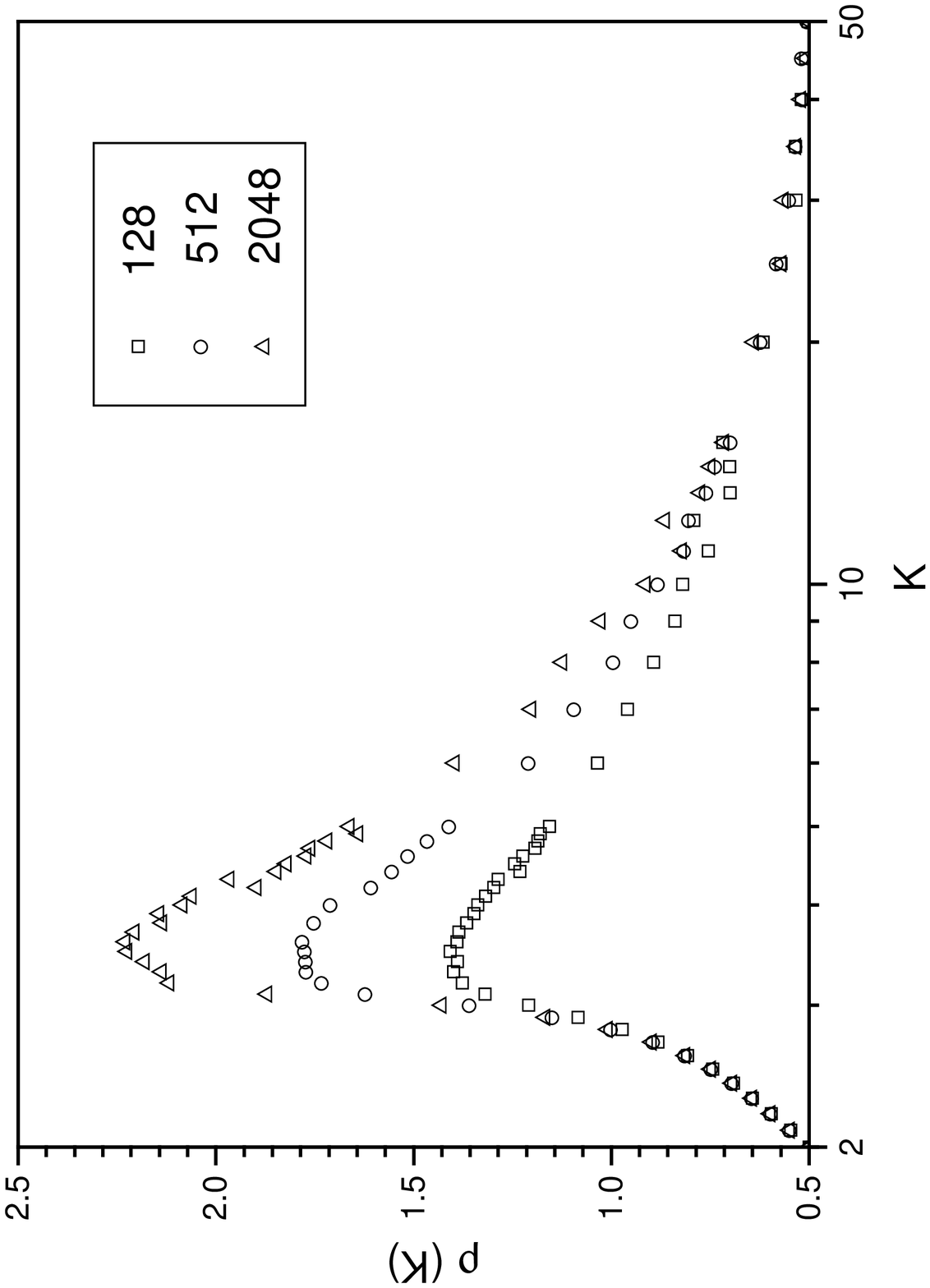}}
\caption{ Long-run performance rate $\rho $ for different values of $k$ and $%
n=128,$ $512,$ $2048.$
}
\end{figure}

Concerning Figure 7, the first observation one may readily make is that, for
any given $k,$ the long-run rate $\rho (k)\equiv \lim_{T\rightarrow \infty
}\rho (T,k)$ is a well-defined magnitude since $\rho (T,k)$ is a (bounded)
monotone function of $T.$ This, however, does not entail any new information
by itself, but is just an indirect confirmation that the avalanche and
advance distributions display long-run distributions with well-defined
averages. For, in view of (\ref{rho}), note that the performance rate can be
simply computed as the ratio of the average advance over the average
avalanche size over a time path.

There are, however, some additional and genuinely new observations arising
from Figures 7 and 8 that will underlie much of our ensuing discussion. For
future reference, it is useful to organize them into the following two
(partially overlapping) conclusions.

\begin{enumerate}
\item[C1]  The system's dynamic performance (as reflected by $\rho )$ enjoys
positive dependence on $n$ only within (or very close to) the critical
region -- cf. Figure 8. This implies that, for large $n,$ the performance
rate $\rho $ is optimized within (or very close to) the critical region.

\item[C2]  The optimal value of $k$ that maximizes $\rho (k)$ is independent
of $n$ and is located at the ``lower edge'' of the critical region, i.e. in
a narrow range around $k=3.5.$
\end{enumerate}

By C1, significant \emph{scale effects} (i.e. improvements in performance
due to population increases) are intimately associated to criticality, i.e.
they only arise when, roughly, $k\in (3,30)$. On the other hand, C2
indicates that the performance of the system is maximized around the point
where such criticality starts to set in. This latter conclusion may be
interpreted along the lines put forward by some authors (e.g. Kauffman
(1993)) who have argued that the dynamic response of large interacting
systems is often optimized at the brink where heterogeneity and ``disorder''
(i.e. criticality in our case) just begin to manifest itself.

The former considerations suggest that our analysis may be provided with
some normative interpretation. Again, to be specific, suppose that $k$ is a
policy variable that determines the ``technological system'' to be applied
in a certain organization (recall our former discussion). Then, if the
underlying conditions are suitably approximated by the model (in particular,
updates are infrequent relative to adjustment and the rate $\rho $ is a
relevant measure of performance), our results indicate that the optimal $k$
is one that induces some ``barely critical'' amount of inter-agent
heterogeneity.

In a related vein, one may instead approach matters from an evolutionary
viewpoint. Specifically, suppose that the underlying environment consists of
a variety of different organizations that are characterized by alternative
technological systems (i.e. idiosyncratic $k$'s), all of them initially
co-existing as part of an heterogenous population. Then, our conclusions
suggest that if evolutionary forces impinge on the population composition
among different types of organizations (say, because of performance-related
differences in survival or through social imitation), the long-run situation
that should eventually prevail is one where organizations are placed at the
edge of criticality.

\section{Theoretical analysis\label{S-TA}}

The entire simulation results reported in Section \ref{NumA} refer to the
numerical scenario given by (\ref{payoffs-sim}) and (\ref{unif}), which is a
particular case of the general context defined by (\ref{payoffs}) and (\ref
{incomp})-(\ref{incomp3}). Similar regularities (i.e., in particular, P1-P3
and C1-C2) arise in alternative scenarios consistent with these conditions.
Given the vast complexity of the underlying process, it would have been very
hard to establish those results by other than numerical methods. However,
once established, we are now in a much better position to build partially
upon them to aim at understanding their relationship analytically.

Specifically, our theoretical approach will start by \emph{postulating}
P1-P3, i.e. it will assume that the system is critical for a certain
parameter range. Then, based on this assumption, we shall strive to \emph{%
derive} analytically the conclusions C1-C2. Overall, this theoretical
exercise will improve our understanding of the important relationship
between criticality and the system's dynamic performance.\bigskip

Our first task is to provide an explicit expression for the performance rate
$\rho .$ For simplicity, we postulate that, for any given $k>0,$ the system
is either in a critical, a supercritical, or a subcritical regime. In
practice, of course, there must be a gradual transition from
supercriticality to criticality and then subcriticality as $k$ grows in the
range $(0,\infty ).$ For our present analytical purposes, however, that
range is assumed \emph{partitioned} into three subregions by thresholds $%
\underline{k}$ and$\,\bar{k}.$ Specifically, the interval $(0,\underline{k})$
is taken to define the supercritical subregion, the interval $(\underline{k},%
\bar{k})$ the critical subregion, and the remaining $(\bar{k},\infty )$ the
subcritical one.

The determination of $\rho $ outside of the critical region is
straightforward. On the one hand, when the system is supercritical $(k<%
\underline{k}),$ avalanches are system-wide (i.e. $s=n)$ and hence $\rho $
is equal to $\bar{\sigma}\equiv \mathbf{E(}\tilde{\sigma}).$ On the other
hand, in the subcritical region $(k>\bar{k}),$ all avalanches can be taken
to be of (approximately) unit size and therefore the rate $\rho $ is again
equal to $\bar{\sigma}.$

Now, consider the critical region where $\underline{k}<k<\bar{k}.$ Denoting
by $P(H)$ and $P(s)$ the corresponding discrete empirical densities, let
\begin{eqnarray*}
\bar{H} &\equiv &\int H\,P(H)\,dH=\lim_{T\rightarrow \infty }\frac{1}{T}%
\sum_{t=1}^{T}H(t) \\
\bar{s} &\equiv &\int s\,P(s)\,ds=\lim_{T\rightarrow \infty }\frac{1}{T}%
\sum_{t=1}^{T}s(t)
\end{eqnarray*}
define the long-run averages of $H$ and $s$. Within the critical region,
these densities are well-defined. Therefore, from (\ref{rho}), we may
compute the performance rate as follows:
\begin{equation}
\rho =\lim_{T\rightarrow \infty }\frac{\bar{H}\cdot T}{\bar{s}\cdot T}=\frac{%
\bar{H}}{\bar{s}}.  \label{rho2}
\end{equation}
Then, we may rely on P1 and P2 to conduct the following (approximate)
derivations:\footnote{%
Of course, this computation must be viewed only as an approximation
conducted under the implicit assumption that the support of the conditional
distributions $P(H\mid s=\bar{s})$ displays a relatively narrow support --
cf. Jensen (1998, p. 38).}
\[
\rho \sim \frac{\int_{1}^{n}s^{\alpha }\,s^{-\gamma }ds}{\int_{1}^{n}s\,s^{-%
\gamma }ds}
\]
which is easily seen to lead to
\begin{equation}
\rho \sim \frac{2-\gamma }{\alpha -\gamma +1}\frac{n^{1+\alpha -\gamma }-1}{%
n^{2-\gamma }-1}.  \label{rhodegamma}
\end{equation}
Clearly, for large $n,$ the above value for $\rho $ is crucially dependent
on the exponents of $n$ (in particular, on whether they are positive or
negative). In this respect, three different cases may be considered, with
respective values for $\rho $ that may be approximated (for large $n)$ as
follows:\footnote{%
Close to the borders between the different regimes, limits must be handled
with care since divergencies pertaining vanishing terms occur and different
limit operations do not commute. These technical issues notwithstanding, a
detailed analysis of (\ref{rhodegamma}) will convince the reader that $\rho $
displays the properties later required (e.g. given $n$ and $\alpha >1,$ it
induces a well-defined decreasing function of $\gamma ).$}

\begin{enumerate}
\item[(i)]  If $\gamma <2$,
\[
\rho \sim \frac{2-\gamma }{\alpha -\gamma +1}n^{\alpha -1};
\]

\item[(ii)]  If $2<\gamma <\alpha +1$,
\[
\rho \sim \frac{\gamma -2}{\alpha -\gamma +1}n^{(\alpha -1)-(\gamma -2)};
\]

\item[(iii)]  If $\gamma >\alpha +1,$
\[
\rho \sim \frac{\gamma -2}{\gamma -\alpha -1}.
\]
\end{enumerate}

By P1-P3, the parameters involved in the above expressions (in particular, $%
\gamma $ and $\alpha )$ can be taken to be independent of $n,$ the
population size. Therefore, for large systems (large $n)$, it follows that
the performance rate $\rho (k)$ will be maximized somewhere in the critical
region, i.e. for $k\in (\underline{k},\bar{k}).$ For only in this region
does the performance of the system benefit from ``scale effects''. Outside
of it (that is, in either the supercritical or subcritical regions), the
performance rate is (approximately) equal to $\bar{\sigma},$ independently
of $n$.

The previous considerations provide the basis for understanding C1. To
understand C2, we now rely on P3, i.e. avalanche sizes and the induced
advances are related through a power law that is not only independent of $n$
but also holds unchanged for all $k$ (of course, as long as $k$ remains in
the critical region). This key feature of critical behavior has the
following striking implication: the effect of $k$ on $\rho $ becomes \emph{%
solely }channelled through its effect on $\gamma $ (i.e. the steepness of
the size distribution). Building upon this key observation, we are now in a
position to understand why the maximization of $\rho $ must indeed be
achieved at the lower edge of the critical region.

To this end, restrict attention to the critical region, and conceive $\rho $
as a function of $\alpha $ and $\gamma ,$ that is extended continuously from
the interior of the different subregions (i)-(iii) for given $n$. Then, for
small changes in $k,$ we can symbolically write:
\begin{eqnarray*}
\frac{\Delta \rho }{\Delta k} &=&\frac{\Delta \rho }{\Delta \gamma }\frac{%
\Delta \gamma }{\Delta k}+\frac{\Delta \rho }{\Delta \alpha }\frac{\Delta
\alpha }{\Delta k} \\
&=&\frac{\Delta \rho }{\Delta \gamma }\frac{\Delta \gamma }{\Delta k}
\end{eqnarray*}
since, by P3, we have $\frac{\Delta \alpha }{\Delta k}=0$ (i.e. $\alpha $ is
unaffected by $k$ within the critical region). Now, it is easy to check from
(\ref{rhodegamma}) that $\frac{\Delta \rho }{\Delta \gamma }<0.$ Combining
this latter fact with $\frac{\Delta \gamma }{\Delta k}>0$ (i.e. the weight
of small avalanches grows with $k$), it follows that $\frac{\Delta \rho }{%
\Delta k}<0$, i.e. $\rho $ is a decreasing function of $k.$ Obviously, this
implies that the maximization of $\rho $ must occur at the lower boundary of
the critical region. As desired, therefore, this provides an (approximate)
analytical basis for those features of our numerical simulations which were
stated in C2.

In a sense, C2 may be interpreted as suggesting that optimal performance
builds upon a rather delicate compromise between ``order'' (synchronization
or homogeneity, i.e. supercriticality) and ``disorder'' (criticality).
Heuristically, the underlying intuition for why \emph{some }such balance
should be expected to arise is not difficult to understand. On the one hand,
if all individuals were to advance in step because of low incompatibility
costs, individual adjustments would always be relatively small and no steep
gradients could ever arise. Consequently, the overall pace of advance should
be slow, individuals hardly taking advantage of the ``scale economies'' that
a large system would avail. But, on the other hand, if incompatibility costs
were large, the scale effects impinging on overall advance that could be
potentially afforded by a large system would be, again, not fully taken
advantage of. In this second case, avalanches would typically be too small
for any heterogeneity to be profited by its required complement: an
effective process of diffusion.

The previous considerations provide quite a clear intuition for why optimal
performance should require criticality, i.e. a suitable trade-off between
homogeneity and heterogeneity in agents' unfolding behavior. What seems
substantially more subtle is the additional sharper conclusion that
optimality is to be expected at the lower edge of the critical region. As
explained, this appears to be intimately associated to the intriguing
``empirical'' (i.e. numerical) finding that criticality induces a fixed
power relationship between avalanche sizes and the induced advances.

\section{Summary and Conclusion}

This paper represents a first step in a research project where we plan to
study in detail the relationship among complexity, optimality, and
self-organization in systems composed by a large number of locally
interacting entities. As suggested above, many social and economic systems
may be modelled in this fashion. In particular, this approach seems
particularly well suited to study technological evolution when the decisions
adopted by the different entities (individual agents, firms, or even
sectors) display local complementarities.

In the context of a simple model with these features, we have seen that if
dis-coordination costs exceed a certain threshold (but are not too large),
the system self-organizes itself into a critical state and the sizes of
diffusion waves are distributed according to a power law. In this case,
moreover, the roughness of the population profile converges to a
well-defined (average) magnitude, thus indicating that there is a specific
degree of long-run population heterogeneity associated to each parameter
(cost) configuration. In fact, it turns out that this ``critical''
heterogeneity plays a crucial role in the performance of the system.
Specifically, we have seen that, given a natural measure of performance
(that admits an interpretation of either average adoption cost or rate of
technological advance), the system behaves optimally within the critical
region -- or, more precisely, at the lower edge of this region. This
suggests that inter-agent heterogeneity plays a crucial role in the
evolution of the system. In other words, either too ``orderly''
(synchronized) or too ``chaotic'' (dis-coordinated) dynamics is detrimental
to performance in that it imposes too frequent (and thus costly) or too rare
(and therefore unduly staggered) adjustment on agents' actions

In ongoing research, we are studying a number of extensions of the present
model. One of them concerns higher dimensional setups (specifically, a
two-dimensional torus), where each individual has more than two neighbors
and therefore the diffusion paths may exhibit richer geometries. A second
extension involves studying irregular (but fixed)\footnote{%
Note that, in order to preserve the key local structure of interaction that
underlies our analysis, the network cannot change in an unrestricted (say,
time-independent) fashion.} networks of the small-world variety (see Watts
and Strogatz (1998)). In each of these contexts, the same qualitative
conclusions found in the present model are essentially maintained. However,
in the latter case (small-world networks), one obtains the expected result
that, due to a relatively short expected path between any two agents, the
parameter range where synchronous behavior tends to arise becomes
significantly larger than in regular networks.

This paper has shown that a quite simple model of social, but local-based,
interaction may produce persistent and wide heterogeneity in the induced
population dynamics. It is clear, however, that not all social networks can
be expected to display such a behavior. For example, complete networks,
which display no local structure, can only exhibit uninteresting ``waves''
in a coordination context analogous to that considered here. In such a
context, fixed-size avalanches of the order of system size (and only those)
would occur as the number of updates accumulated since the last avalanche
come to exceed a certain threshold.\footnote{%
For simplicity, we are implicitly assuming every update to be of the same
\emph{given }magnitude.}

Naturally, the above point raises the question of what networks might be
conducive to critical behavior and, more importantly, whether such a class
of criticality-supporting networks could arise \emph{endogenously} when the
social network is not fixed but may also co-evolve as dictated by agents'
own decisions. The issue, in a sense, is analogous to that addressed by the
most recent developments of evolutionary game theory that focus on how the
parallel co-evolution of \emph{both} players' links (or connections) and
their actions may affect the received analysis on equilibrium selection.%
\footnote{%
See, for example, Bala and Goyal (2000), Jackson and Watts (2000), or Goyal
and Vega-Redondo (2000).} Indeed, not only the concerns but also some of the
techniques used in this literature (e.g. those employed to analyze perturbed
Markov processes) would seem quite applicable to the problem at hand. The
study of this important topic is left for future research.

\bigskip \pagebreak

{\Large References\medskip }

Aghion, P. and P. Howitt, 1992. A model of growth through creative
destruction'', \emph{Econometrica }\textbf{60}, 323-51.{\Large \medskip }

Agliardi, E., 1998. \emph{Positive Feedback Economies}, London: MacMillan.%
{\Large \medskip }

Albert, R., H. Jeong, A.-L. Barabasi, 1999. Diameter of the World-Wide
Web'', \emph{Nature }\textbf{401}, 130-31.{\Large \medskip }

Arenas, A., C. P\'{e}rez, A. D\'{i}az-Guilera, and F. Vega-Redondo, 2000.
Self-organized evolution in socio-economic environments'' \emph{Physical
Review E} \textbf{61}, 3466-69.{\Large \medskip }

Arthur, W.B., S.N. Durlauf \& D.A. Lane 1997. \emph{The Economy as an
Evolving System, }Santa Fe Institute, Proceedings Volume XXVII, Reading,
Mass.: Addison-Wesley.{\Large \medskip }

Bak, P. C. Tang, and K. Wiesenfeld, 1987. Self-organized criticality: an
explanation of $1/f$ noise'', \emph{Phys. Rev. Lett. }\textbf{59}, 381.%
{\Large \medskip }

Bak, P., C. Tang, and K. Wiesenfeld, 1988. Self-organized criticality'',
\emph{Phys. Rev. A }\textbf{38}, 364.{\Large \medskip }

Bak, P. 1996. \emph{How Nature Works: The Science of Self-Organized
Criticality}, New York: Copernicus.{\Large \medskip }

Bala, V. and S. Goyal, 2000. A Non-Cooperative Model of Network Formation'',
\textit{Econometrica}, \textbf{68}, 5,{\Large \medskip } 1181-1231.{\Large %
\medskip }

Barabasi A.-L. and H.E. Stanley, 1995. \emph{Fractal Concepts in Surface
Growth}, Cambridge: Cambridge University Press.{\Large \medskip }

Blume, L., 1993. The statistical mechanics of strategic interaction'', \emph{%
Games and Economic Behavior }\textbf{5}, 387-424.{\Large \medskip }

Bryant, J., 1983. A simple rational expectations Keynes type model'', \emph{%
Quarterly Journal of Economics }\textbf{98}, 525-529.{\Large \medskip }

Crawford, V., 1991. An evolutionary interpretation of van Huyck, Battalio,
and Beil's experimental results on coordination'', \emph{Games and Economic
Behavior }\textbf{3}, 25-59.{\Large \medskip }

Economides, N., 1996. The economics of networks'', \emph{International
Journal of Industrial Organization} \textbf{16}, 673-699.{\Large \medskip }

Ellison, G., 1993. Learning, local interaction, and coordination'', \emph{%
Econometrica} \textbf{4}, 1047--73.{\Large \medskip }

Faloutsos, M., P. Faloutsos, and C. Faloutsos, 1999. On power-law
relationships of the internet topology'', ACM SIGCOMM '99, \emph{Comput.
Commun. Rev. }\textbf{29}, 251-63.{\Large \medskip }

Farrel, J. \& J. Saloner, 1985. Standarization, compatibility, and
innovation'', \emph{Rand Journal of Economics }\textbf{16} 70-83.

Goyal, S. and F. Vega-Redondo, 2000. Learning, Network Formation and
Coordination'', mimeo, Erasmus University (Rotterdam) and Universidad de
Alicante.{\Large \medskip }

Grossman, G.M. and E. Helpman, 1991. Quality ladders in the theory of
growth'', \emph{Review of Economic Studies }\textbf{58}, 43-61.{\Large %
\medskip }

Huberman, B.A. and L.A. Adamic, 1999. Growth dynamics of the World-Wide
Web'', \emph{Nature }\textbf{401}, 131.{\Large \medskip }

Jackson, M. and A. Watts, 2000. On the formation of interaction networks in
social coordination games'', mimeo, Caltech.{\Large \medskip }

Jensen, H. J., 1998. \emph{Self-Organized Criticality}, Cambridge: Cambridge
University Press.{\Large \medskip }

Kandori, M., G. Mailath \& R. Rob, 1993. Learning, mutations, and long-run
equilibria in games'', \emph{Econometrica}\textbf{\ 61}, 29-56.{\Large %
\medskip }

Kaufman, S., 1993. \emph{The Origins of Order: Self Organization and
Selection in Evolution, }Oxford: Oxford University Press.{\Large \medskip }

Katz, M. and C. Shapiro, 1985. Network externalities, competition, and
compatibility'', \emph{American Economic Review }\textbf{75}, 424-40.{\Large %
\medskip }

Krugman, P., 1996.\ \emph{The self-organizing economy}, Cambridge Mass.:
Blackwells.

Redner, S. 1998. How popular is your paper: an empirical study of the
citation distribution'', \emph{European Phys. J. B }\textbf{4}, 131-34.%
{\Large \medskip }

Scheinkman, J. and M. Woodford, 1994. Self-organized criticality and
economic fluctuations'', \emph{American Economic Review }\textbf{84},
417-421.{\Large \medskip }

Sergestrom, P.S., T.C.A. Anant, and E. Dinopoulos, 1990. A Schumpeterian
model of the product life cycle'', \emph{American Economic Review }\textbf{80%
}, 1077-91.{\Large \medskip }

van Huyck, J.B., R. Battalio, and R.O. Beil, 1990. Tacit coordination games,
strategic uncertainty, and coordination failure'', \emph{American Economic
Review }\textbf{80}, 234-48.{\Large \medskip }

Watts, D.J. and S.H. Strogatz, 1998. Collective dynamics of `small-world'
networks'', \emph{Nature }\textbf{393}, 440-42.{\Large \medskip }

Young, P., 1993. The evolution of conventions'', \emph{Econometrica }\textbf{%
61}, 57-84.{\Large \medskip }

Young, P., 1998. Diffusion in social networks'', John Hopkins University,
mimeo.

\end{document}